# Interfacial and Confined Optical Phonons in Wurtzite Nanocrystals

Vladimir A. Fonoberov[*] and Alexander A. Balandin[†]

*Nano-Device Laboratory, Department of Electrical Engineering*
*University of California–Riverside, Riverside, California 92521*

We derive within the dielectric-continuum model an integral equation that defines interface and confined polar optical-phonon modes in nanocrystals with wurtzite crystal structure. It is demonstrated theoretically, that while the frequency of confined polar optical phonons in zincblende nanocrystals is equal to that of the bulk crystal phonons, the confined polar optical phonons in wurtzite nanocrystals have a discrete spectrum of frequencies different from those of the bulk crystal. The calculated frequencies of confined polar optical phonons in wurtzite ZnO nanocrystals are found to be in an excellent agreement with the experimental resonant Raman scattering spectra of spherical ZnO quantum dots.

Optical lattice vibrations (phonons) manifest themselves in Raman, absorption, and photoluminescence spectra of bulk crystals and nanocrystalline materials. In this sense, optical phonon frequency is a unique signature for a given crystalline material. However, the phonon spectra of nanocrystals can be very different from those of bulk crystals, and depend on the nanocrystal shape and size. The latter explains an importance of developing theoretical (preferably – analytical) tools for calculation of phonon modes in nanocrystals of different shapes and crystal structures. About four decades ago, Englman and Ruppin [1] found that in zincblende nanocrystals there exist confined phonon modes with the frequencies equal to those of bulk transverse optical (TO) and longitudinal optical (LO) phonons. It was also established that the interface phonon modes in such nanocrystals have frequencies intermediate between those of LO and TO modes. Later, Knipp and Reinecke [2] developed an approach to calculate the interface optical phonon modes in the isotropic quantum dots of arbitrary shapes. A successful theoretical explanation of experimental Raman, absorption, and photoluminescence spectra has been obtained for a variety of optically isotropic nanocrystals with different shapes [3-6].

Very recently, wide-bandgap wurtzite nanocrystals such as ZnO and GaN quantum dots have attracted attention as promising candidates for optoelectronic, electronic and biomedical applications. The hexagonal wurtzite (space group $C_{6v}^4$) nanocrystals have uniaxial crystal structure with the optical axis, which coincides with the *c*-axis of the crystal. Due to this uniaxial anisotropy, the confined optical and interface phonon modes in wurtzite quantum dots can be substantially different from those in zincblende (isotropic) quantum dots. In a view of this difference and expected applications, it is very important and timely to investigate optical phonons in optically anisotropic wurtzite nanocrystals. A recent theoretical study [7] of optical phonons in a two-dimensional system – wurtzite GaN/AlN quantum well – has shown that the intrinsic anisotropy of wurtzite material results in the qualitative difference of the phonon spectrum from that for optically isotropic quantum wells.

---

[*] Electronic address (V.A. Fonoberov): vladimir@ee.ucr.edu
[†] Electronic address (A.A. Balandin): alexb@ee.ucr.edu

In this Letter we develop an analytical approach to calculate polar optical-phonon modes in wurtzite quantum dots of arbitrary shapes. Using the developed approach, we find the polar optical-phonon modes in spheroidal optically anisotropic quantum dots. To the best of our knowledge, this it the first report of the solution to this problem. The calculated discrete frequencies of polar optical phonons in ZnO nanocrystals are in excellent agreement with experimental resonant Raman scattering spectra of strain-free spherical ZnO quantum dots [8].

Let us consider an arbitrary shaped nanocrystal with a uniaxial anisotropy of the crystal lattice. The exterior medium is also assumed to be uniaxially anisotropic with the same direction of the symmetry axis ($z$-axis) as that for the nanocrystal. Within the framework of the dielectric-continuum approximation and Loudon's models for uniaxial crystals, the dielectric tensors in the nanocrystal ($k = 1$) and exterior medium ($k = 2$) can be written as

$$\hat{\varepsilon}^{(k)}(\omega) = \begin{pmatrix} \varepsilon_{\perp}^{(k)}(\omega) & 0 & 0 \\ 0 & \varepsilon_{\perp}^{(k)}(\omega) & 0 \\ 0 & 0 & \varepsilon_{z}^{(k)}(\omega) \end{pmatrix}, \qquad (1)$$

$$\varepsilon_{\perp}^{(k)}(\omega) = \varepsilon_{\perp}^{(k)}(\infty)\frac{\omega^2 - \left(\omega_{\perp,\text{LO}}^{(k)}\right)^2}{\omega^2 - \left(\omega_{\perp,\text{TO}}^{(k)}\right)^2}; \quad \varepsilon_{z}^{(k)}(\omega) = \varepsilon_{z}^{(k)}(\infty)\frac{\omega^2 - \left(\omega_{z,\text{LO}}^{(k)}\right)^2}{\omega^2 - \left(\omega_{z,\text{TO}}^{(k)}\right)^2}, \qquad (2)$$

where $\varepsilon_{\perp}(\infty)$ and $\varepsilon_{z}(\infty)$ are optical dielectric constants, $\omega_{\perp,\text{LO}}$ and $\omega_{z,\text{LO}}$ are LO phonon frequencies, and $\omega_{\perp,\text{TO}}$ and $\omega_{z,\text{TO}}$ are TO phonon frequencies of the corresponding bulk materials. The phonon potential $V(\mathbf{r})$ of the polar optical modes satisfies the Maxwell's equation

$$-\nabla\left(\hat{\varepsilon}^{(k)}(\omega)\nabla V_k(\mathbf{r})\right) = 0 \quad (k=1,2) \qquad (3)$$

with the boundary conditions

$$V_1(\mathbf{A}) = V_2(\mathbf{A}); \quad D_1(\mathbf{A}) = D_2(\mathbf{A}), \qquad (4)$$

where $\mathbf{A} \in S$ is the radius-vector of the interface $S$ between the media 1 and 2 and

$$D_k(\mathbf{A}) = \mathbf{n}_A \hat{\varepsilon}^{(k)}(\omega)\left(\nabla V_k(\mathbf{r})\right)\Big|_{\mathbf{r}=\mathbf{A}} \qquad (5)$$

is the projection of the displacement vector $\mathbf{D}$ on the outer normal $\mathbf{n}_A$ at the point $\mathbf{A}$ of the interface.

To solve Eq. (3) with boundary conditions (4), let us first find the Green's function $G_k(\mathbf{r},\mathbf{r}';\omega)$ for each medium as a solution of the equation

$$-\nabla\left(\hat{\varepsilon}^{(k)}(\omega)\nabla G_k(\mathbf{r},\mathbf{r}';\omega)\right) = \delta(\mathbf{r}-\mathbf{r}'). \qquad (6)$$

It can be easily shown that $G_k(\mathbf{r},\mathbf{r}';\omega)$ has a simple analytical form:

$$G_k(\mathbf{r},\mathbf{r}';\omega) = \frac{1}{4\pi\varepsilon_{\perp}^{(k)}(\omega)}\left(g^{(k)}(\omega)(\boldsymbol{\rho}-\boldsymbol{\rho}')^2 + (z-z')^2\right)^{-1/2}, \qquad (7)$$

where $\boldsymbol{\rho}$ is the radius-vector in the $xy$-plane and

$$g^{(k)}(\omega) = \varepsilon_{z}^{(k)}(\omega)/\varepsilon_{\perp}^{(k)}(\omega). \qquad (8)$$

Now, integrating the difference of Eq. (6) multiplied by $V_k(\mathbf{r})$ and Eq. (3) multiplied by $G_k(\mathbf{r},\mathbf{r}';\omega)$ over the nanocrystal's volume $\Omega_1$ for $k = 1$ and over the exterior medium's volume $\Omega_2$ for $k = 2$, then applying the Green's theorem to both integrals, we obtain

$$V_k(\mathbf{r}')\eta_k(\mathbf{r}') = (-1)^{k-1}\int_S \left(D_k(\mathbf{A})G_k(\mathbf{A},\mathbf{r}';\omega) - V_k(\mathbf{A})\mathbf{n}_\mathbf{A}\mathbf{F}_k(\mathbf{A},\mathbf{r}';\omega)\right)dA, \qquad (9)$$

where $\eta_k(\mathbf{r}')$ is equal to 1/2 when $\mathbf{r}' \in S$ and to 1 in all other points of the volume $\Omega_k$ and

$$\mathbf{F}_k(\mathbf{r},\mathbf{r}';\omega) = -\frac{\mathbf{r}-\mathbf{r}'}{4\pi}g^{(k)}(\omega)\left(g^{(k)}(\omega)(\boldsymbol{\rho}-\boldsymbol{\rho}')^2 + (z-z')^2\right)^{-3/2}. \qquad (10)$$

Taking $\mathbf{r}' = \mathbf{A}'$ in Eq. (9) and using boundary conditions (4), we arrive at the following homogeneous system of two integral equations

$$\begin{cases} \int_S \left(D(\mathbf{A})G_1(\mathbf{A},\mathbf{A}';\omega) + V(\mathbf{A})\left[-\mathbf{n}_\mathbf{A}\mathbf{F}_1(\mathbf{A},\mathbf{A}';\omega) - \tfrac{1}{2}\delta(\mathbf{A}-\mathbf{A}')\right]\right)dA = 0, \\ \int_S \left(D(\mathbf{A})G_2(\mathbf{A},\mathbf{A}';\omega) + V(\mathbf{A})\left[-\mathbf{n}_\mathbf{A}\mathbf{F}_2(\mathbf{A},\mathbf{A}';\omega) + \tfrac{1}{2}\delta(\mathbf{A}-\mathbf{A}')\right]\right)dA = 0. \end{cases} \qquad (11)$$

Frequencies $\omega$ that allow for a nonzero solution of Eq. (11) define the spectrum of polar optical phonons in the nanocrystal, while the corresponding pairs $\{D(\mathbf{A}), V(\mathbf{A})\}$ define the projection of the displacement vector on the outer normal and the phonon potential at the interface $S$. The latter, in its turn, defines the phonon potential in the whole space, according to Eq. (9).

Thus, we have derived an equation that defines both interface and confined polar optical-phonon modes in quantum dots with wurtzite crystal structure. It can be shown that for zinc-blende quantum dots, where $\varepsilon_\perp(\omega) = \varepsilon_z(\omega)$, Eq. (11) reduces to the one obtained by Knipp and Reinecke [2] for interface optical phonons. It is seen from Eqs. (7) and (10) that kernels $G_k(\mathbf{A},\mathbf{A}';\omega)$ and $-\mathbf{n}_\mathbf{A}\mathbf{F}_k(\mathbf{A},\mathbf{A}';\omega)$ of Eq. (11) are singular when $\mathbf{A} = \mathbf{A}'$, but this singularity is integrable for an interface $S$ that does not contain sharp edges or cusps where the normal vector $\mathbf{n}_\mathbf{A}$ is not defined. Nanocrystals with realistic parameters do not have such edges or cusps. Thus, one can conclude that the spectrum of eigenfrequencies of Eq. (11) is discrete [9]. This conclusion corrects the suggestion of Ref. [10] of the existence of continuous allowed phonon frequency windows in anisotropic nanocrystals.

The numerical solution of Eq. (11) for a wurtzite nanocrystal of an arbitrary shape can be found using standard techniques. However, an analytical solution would be much more convenient for practical applications. Therefore, let us consider a uniaxial ellipsoid with the symmetry axis $z$ and the ratio of semi-axes $\gamma = c/a$. The shape of most of the practically important wurtzite nanocrystals can be modeled as ellipsoidal. For example, colloidal nanocrystals have nearly spherical ($\gamma = 1$) shape; nanorods and epitaxial quantum dots can be modeled as prolate ($\gamma > 1$) and oblate ($\gamma < 1$) ellipsoids, respectively. Introducing new coordinates with $z' = z/\gamma$, the interface $S$ becomes a sphere of radius $a$, which can be described with spherical angles $\theta$ and $\phi$. It can be shown that the solution of Eq. (11) in the new coordinates is

$$V(\theta,\phi) = Y_{l,m}(\theta,\phi); \quad D(\theta,\phi) = B Y_{l,m}(\theta,\phi), \qquad (12)$$

where $Y_{l,m}(\theta,\phi)$ are spherical harmonics. Substituting Eq. (12) into Eq. (11), we obtain after the integration:

$$c\gamma B = \varepsilon_z^{(1)}(\omega)\left(\xi\frac{d\ln P_l^m(\xi)}{d\xi}\right)\bigg|_{\xi=(1-g^{(1)}(\omega)/\gamma^2)^{-1/2}} = \varepsilon_z^{(2)}(\omega)\left(\xi\frac{d\ln Q_l^m(\xi)}{d\xi}\right)\bigg|_{\xi=(1-g^{(2)}(\omega)/\gamma^2)^{-1/2}}, \qquad (13)$$

where $P_l^m$ and $Q_l^m$ are associated Legendre functions of the first and second kind, respectively. Roots of Eq. (13) define the spectrum of polar optical phonons in spheroidal nanocrystals.

Recently measured Raman spectra of spherical wurtzite ZnO nanocrystals in air [8] have revealed confined polar optical phonons that have frequencies different from those of bulk LO and TO phonons and lying in the intervals ($\omega_{z,\text{LO}}$, $\omega_{\perp,\text{LO}}$) and ($\omega_{z,\text{TO}}$, $\omega_{\perp,\text{TO}}$). No quantitative explanation of the observed frequencies of the confined phonons has been given so far. In the following, we will calculate the polar optical-phonon modes in spherical ZnO quantum dots ($\gamma = 1$) and compare them with the reported Raman spectra [8]. For an optically inactive exterior medium, such as air, plastic, glass, etc., the dielectric tensor reduces to a constant $\hat{\varepsilon}^{(2)}(\omega) = \varepsilon_D$. In this case $g^{(2)}(\omega) = 1$ and we can explicitly write Eq. (13) in the following algebraic form:

$$\sum_{n=0}^{\lfloor \frac{l-|m|}{2} \rfloor} \binom{l-|m|}{2n} \frac{(2n-1)!!(2l-2n-1)!!}{(2l-1)!!} \left( \frac{\varepsilon_\perp^{(1)}(\omega)}{\varepsilon_D} |m| + \frac{\varepsilon_z^{(1)}(\omega)}{\varepsilon_D}(l-|m|-2n) + l + 1 \right) \left( \frac{\varepsilon_z^{(1)}(\omega)}{\varepsilon_\perp^{(1)}(\omega)} - 1 \right)^n = 0. \quad (14)$$

Solutions of Eq. (14) for each pair of integer quantum numbers $l = 1, 2, 3, \ldots$ and $m$ ($-l \leq m \leq l$) give the spectrum of both interface and confined polar optical phonons for the spherical quantum dot. Note, that there are no phonons with $l = 0$ and all phonon frequencies with $m \neq 0$ are twice degenerate with respect to the sign of $m$. It should be pointed out, that in the case of the isotropic spherical nanocrystal [$\varepsilon_\perp(\omega) = \varepsilon_z(\omega)$], Eq. (14) reduces to the equation $\varepsilon(\omega)/\varepsilon_D = -1 - 1/l$ [1] for interface optical phonons. Analyzing Eq. (14), one can find that, unlike spherical zincblende nanocrystals, the spectrum of interface and confined optical phonons in wurtzite nanocrystals depends on $m$. Moreover, for each pair ($l, m$) there is one interface optical phonon and $l - |m|$ confined optical phonons for $m \neq 0$ ($l - 1$ for $m = 0$).

Fig. 1 shows the calculated spectrum of polar optical phonons with $l = 1, 2, 3, 4$ and $m = 0$ in a spherical wurtzite ZnO quantum dot as a function of the optical dielectric constant of the exterior medium $\varepsilon_D$. All optical parameters of wurtzite ZnO are taken from Ref. [11]. The spectrum in Fig. 1 is divided into three regions: confined TO phonons ($\omega_{z,\text{TO}} < \omega < \omega_{\perp,\text{TO}}$), interface phonons ($\omega_{\perp,\text{TO}} < \omega < \omega_{z,\text{LO}}$), and confined LO phonons ($\omega_{z,\text{LO}} < \omega < \omega_{\perp,\text{LO}}$). The above division into confined and interface phonons is based on the sign of the function $g(\omega)$ [see Eq. (8)]. We call the phonons with eigenfrequency $\omega$ interface phonons if $g(\omega) > 0$ and confined phonons if $g(\omega) < 0$. It is seen from Fig. 1 that the frequencies of interface optical phonons decrease substantially when $\varepsilon_D$ changes from the vacuum's value ($\varepsilon_D = 1$) to the ZnO nanocrystal's value ($\varepsilon_D = 3.7$). At the same time the frequencies of confined optical phonons decrease only slightly with $\varepsilon_D$.

Raman spectra of two samples of spherical wurtzite ZnO quantum dots were recorded at room temperature in Ref. [8]. The two samples contained the powder of ZnO nanocrystals with diameters 8.5 nm and 4.0 nm, correspondingly. Since the excitation wavelength 363.8 nm (~ 3.41 eV) was very close to the exciton ground state energy for the 8.5 nm quantum dots [12], the spectrum of the first sample is the resonant Raman spectrum. Both spectra exhibited two peaks in the region of polar optical phonons: 393 cm$^{-1}$ and 588 cm$^{-1}$ for the first sample and 393 cm$^{-1}$ and 584 cm$^{-1}$ for the second sample. The spectrum of polar optical phonons in spherical nanocrystals does not depend on the size of the nanocrystal, however a phonon line in the non-resonant Raman spectrum can be shifted due to the contribution from the excited exciton states. Therefore, if our calculations are correct, we should expect that frequencies 393 cm$^{-1}$ and 588 cm$^{-1}$ of the resonant Raman spectrum correspond to some of the discrete eigenfrequencies found from Eq. (14). Indeed, we can see it in Fig. 1 that the above two frequencies are the one of confined TO phonon with $l = 4$, $m = 0$ and the one of confined LO phonon with $l = 2$, $m = 0$ (we take $\varepsilon_D = 1$ because the experimental spectra were recorded in air).

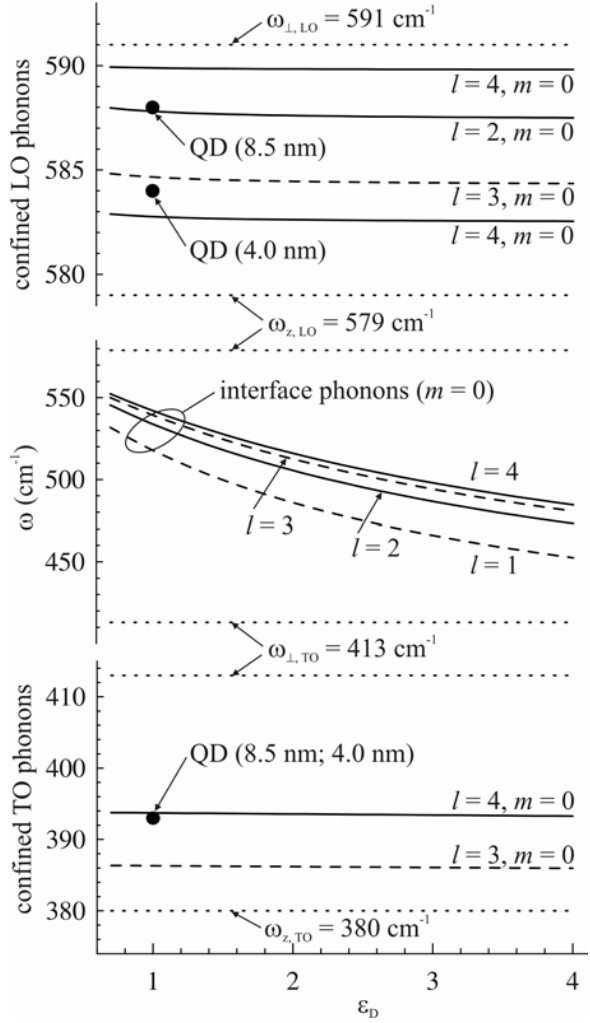

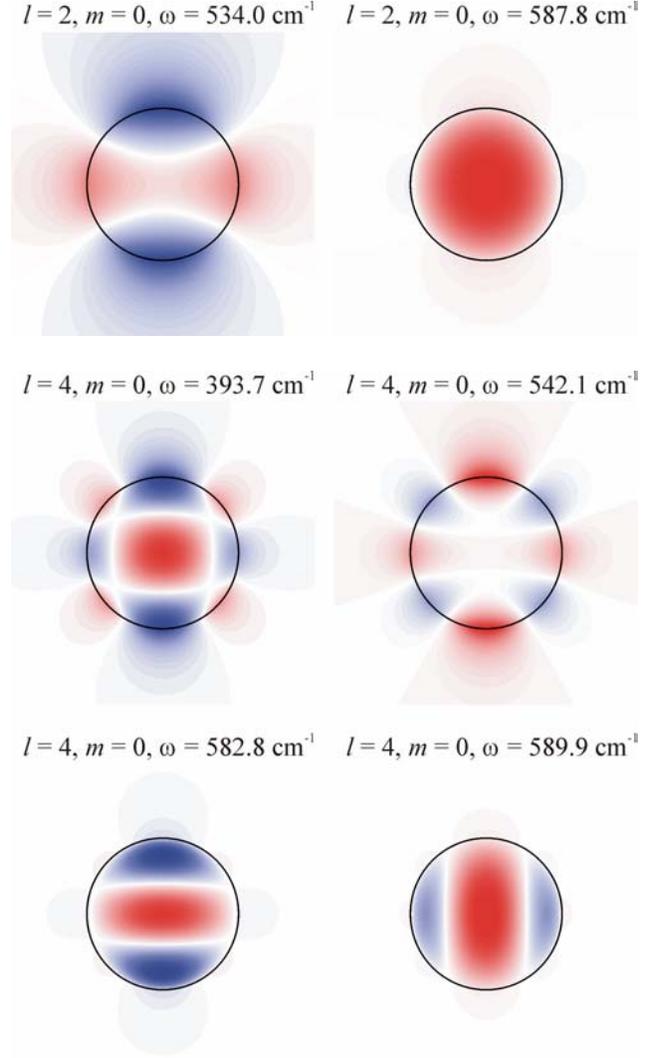

Fig. 1. Spectrum of several polar optical phonon modes in spherical wurtzite ZnO nanocrystals as a function of the optical dielectric constant of the exterior medium. Note that the scale of frequencies is different for confined LO, interface, and confined TO phonons. Large black dots show the experimental points from Ref. [8].

Fig. 2. Cross-sections of the phonon potentials of several polar optical phonon modes in spherical wurtzite ZnO nanocrystals ($\varepsilon_D = 1$). Black circles indicate the boundary of the nanocrystal. The $z$-axis is directed vertically. In the color version, blue and red colors denote positive and negative values of the phonon potential, respectively.

Finally, we have to explain why only the two above phonon modes are active in the resonant Raman spectrum. The dominant component of the wave function of the exciton ground state in spherical ZnO quantum dots is symmetric with respect to the rotations along the $z$-axis or the reflection in the *xy*-plane [12]. Therefore, the selection rules for the optical phonon modes observed in the resonant Raman spectra of ZnO nanocrystals are $m = 0$ and $l = 2, 4, 6, \ldots$ The phonon modes with higher symmetry are more likely to be observed in the Raman spectra. In Fig. 2, the phonon potentials of all two phonon modes with $l = 2$ and all four phonon modes with $l = 4$ are shown. Two interface ($\omega = 534.0$ cm$^{-1}$ and $\omega = 542.1$ cm$^{-1}$) and four confined phonon modes can be seen in Fig. 2. Among the confined phonon modes there is one TO mode ($\omega = 393.7$ cm$^{-1}$) and three LO modes. It is seen from Fig. 2, that the confined LO phonon mode with $l = 2$, $m = 0$ ($\omega = 587.8$ cm$^{-1}$) and the confined TO mode with $l = 4$, $m = 0$ ($\omega = 393.7$ cm$^{-1}$) are the confined phonon modes with the highest symmetry. Therefore, they should give the main contribution to the resonant Raman spectrum. Indeed, they almost coincide with the experimentally recorded frequencies 588 cm$^{-1}$ and 393 cm$^{-1}$.

In conclusion, we derived an integral equation that defines interface and confined polar optical-phonon modes in optically anisotropic wurtzite nanocrystals. It is has been shown that while the frequency of confined polar optical phonons in zincblende nanocrystals is equal to that of the bulk crystal phonons, the confined polar optical phonons in wurtzite nanocrystals have a discrete spectrum of frequencies different from those of the bulk crystal. Based on our analytical solution for spheroidal nanocrystals we were able, for the first time, to explain quantitatively the positions of the polar optical phonon lines observed in the resonant Raman spectra of spherical wurtzite ZnO quantum dots. The proposed model allows one to accurately predict the frequencies of both interface and confined phonons in optically anisotropic nanocrystals.

This work was supported in part by the NSF-NATO 2003 award to V. A. F., ONR Young Investigator Award to A. A. B. and by the DMEA/DARPA CNID program A01809-23103-44.